\documentclass[10pt]{article}
\pdfoutput=1
\usepackage{amssymb,amsmath,mathrsfs,enumerate}
\usepackage{graphicx,rotate,multicol,wrapfig,tabu}
\usepackage[margin=10pt,labelfont=bf, font=it]{caption}
\usepackage{cite}
\usepackage[titletoc,title]{appendix}
\usepackage[colorlinks=true,
            linkcolor=red,
            urlcolor=blue,
            citecolor=blue]{hyperref}
\usepackage{physics}
\usepackage[utf8]{inputenc}

\usepackage[framemethod=default]{mdframed}
\usepackage{showexpl}

\mdfdefinestyle{exampledefault}{%
rightline=true,innerleftmargin=10,innerrightmargin=10,
frametitlerule=true,frametitlerulecolor=green,
frametitlebackgroundcolor=yellow,
frametitlerulewidth=2pt}
\usepackage{color}
\usepackage{braket}
\long\def\rpl#1!!#2!!{\textcolor{red}{#1} \textcolor{blue}{#2}}
\newcommand{\specialcell}[2][c]{%
  \begin{tabular}[#1]{@{}c@{}}#2\end{tabular}}

\makeatletter \renewcommand\d[1]{\ensuremath{%
  \;\mathrm{d}#1\@ifnextchar\d{\!}{}}}
\makeatother

\def\ml{\mathscr L}

\def\nbb{0\nu\beta\beta~{\rm decay}}
\def \order(#1){{\cal O} \left(#1 \right)}
\def\mkpp{m_{\kappa^{++}}}
\def\mcpp{m_{\chi^{++}}}
\def\mcp{m_{\chi^{+}}}

\evensidemargin=\oddsidemargin
\def\Eqn#1{Eq.\ (\ref{#1})}
\def\Eqs#1#2{Eqs.\ (\ref{#1}) and (\ref{#2})}
\allowdisplaybreaks
\begin{document}

\hfill {\small FTUV-17-0104, IFIC-16-97}\\[0.5cm]

\begin{center}
	{\Large \bf A model of neutrino mass and dark matter \\ \vspace*{1.5mm} with large neutrinoless double
	beta decay} \\
	\vspace*{0.8cm} {\sf Julien Alcaide\footnote{julien.alcaide@uv.es}, ~\sf Dipankar	Das\footnote{dipankar.das@uv.es}, ~Arcadi Santamaria\footnote{arcadi.santamaria@uv.es}} \\
	\vspace{10pt} {\small\em Departament de F\'{i}sica T\`{e}orica, Universitat de Val\`{e}ncia and IFIC, Universitat de Val\`{e}ncia-CSIC  \\
	Dr. Moliner 50, E-46100 Burjassot (Val\`{e}ncia), Spain}
	
	\normalsize
\end{center}

\begin{abstract}
We propose a model where neutrino masses are generated at three loop order but neutrinoless double beta decay occurs at one loop. Thus we can have large neutrinoless double beta decay observable in the future experiments even when the neutrino masses are very small. The model receives strong constraints from the neutrino data and lepton flavor violating decays, which substantially reduces the number of free parameters. Our model also opens up the possibility of having several new scalars below the TeV regime, which can be explored at the collider experiments. Additionally, our model also has an unbroken $Z_2$ symmetry which allows us to identify a viable Dark Matter candidate.
\end{abstract}

\bigskip
%===============================================
\section{Introduction}
\label{s:intoduction}
Almost all the extensions of the Standard Model~(SM) directed towards an explanation for the neutrino masses brings in the possibility of
lepton number violation~(LNV) as an outcome. It is well known that neutrinoless double beta
decay~($\nbb$) which is a convincing signature for LNV, will be an inevitable consequence if
the neutrino has Majorana mass. If the main contribution to $\nbb$ proceeds through the Majorana neutrino propagator,
depending on the spectrum of the neutrino masses,
the expected rate for $\nbb$ might be too small to be observed in the experiments. 
But there exist scenarios where the dominant mechanism for $\nbb$ is not controlled
by the Majorana neutrino propagator. In such cases we can have the possibility of large
$\nbb$ even when the neutrino Majorana masses are small. Many studies have been performed
in this direction in the past (see Refs.~\cite{Schechter:1981bd,Mohapatra:1998rq,Vergados:2012xy} for a general overview,
Refs.~\cite{Pati:1974yy,Mohapatra:1974gc,Senjanovic:1975rk,Hirsch:1996qw,Atre:2009rg,Tello:2010am,Blennow:2010th,Ibarra:2010xw,Mitra:2011qr,Dev:2013vxa,Dev:2014xea,Choubey:2012ux,Babu:2010vp,Gustafsson:2012vj,Liu:2016mpf,Jin:2015cla,Okada:2015hia,Ahriche:2014cda,Hatanaka:2014tba,Chen:2014ska,Geng:2015sza,Nishiwaki:2015iqa,Ahriche:2015wha} for specific models\footnote{See also \cite{Helo:2015fba} for a recent review of neutrino mass models in connection to $\nbb$.} and Refs.~\cite{Choi:2002bb,Engel:2003yr,deGouvea:2007qla,delAguila:2012nu} for effective field theory~(EFT) approaches). In Ref.~\cite{delAguila:2012nu}, the authors
performed an EFT analysis of the different ways of generating $\nbb$ and light neutrino
masses by including operators involving only leptons, Higgs and gauge bosons. This led
to a class of interesting models where $\nbb$ was generated at tree level whereas
neutrino masses would appear only at two-loops (see Refs.~\cite{delAguila:2011gr}
for example models in this category).

The model in Ref.~\cite{delAguila:2011gr} contains an $SU(2)_L$ singlet doubly charged
scalar like in the Zee-Babu model\cite{Zee:1985id,Babu:1988ki,Babu:2002uu}, an $SU(2)_L$
triplet scalar with hypercharge $+1$ and a real singlet scalar. A $Z_2$ symmetry, which is
later broken spontaneously, is required to prevent tree-level neutrino masses. The model is
economical in the sense that it contains no new fermions and by design, it gives new
contributions to $\nbb$, which, in principle, can be large. Additionally, it has a rich
phenomenology which can be probed through the searches for the lepton flavor 
violating~(LFV) signals and/or the direct searches for the new scalars in the collider
experiments.

In this article we will present a simple variation of the model in Ref.~\cite{delAguila:2011gr}.
Our new model will have the same field content as in Ref.~\cite{delAguila:2011gr}, except that
the $Z_2$ symmetry will not be broken spontaneously. Consequently, $\nbb$ will now occur
at one-loop whereas neutrino masses will appear at three-loop order. The fact the $Z_2$ is exact makes the model simpler and allows for a viable Dark Matter~(DM) candidate: the lightest of the electrically neutral $Z_2$-odd particles. 
On the other hand, the model keeps all the virtues of the previous model: very predictive neutrino mass matrix, large $\nbb$ decay, rich lepton flavour violation phenomenology and new scalars which are in the sub-TeV region and therefore, are within the reach of the collider experiments in the near future.

Our paper will be organized as follows. In Sec.~\ref{s:model} we lay out the scalar field content and the
physical spectrum of our model. In Sec.~\ref{s:nbb} we discuss the $\nbb$ and the bounds that follow from
it. Neutrino masses and constraints from LFV decays are discussed in Sec.~\ref{sec:nu-mass} and
Sec.~\ref{s:LFV} respectively. We analyze the feasibility of DM in Sec.~\ref{s:DM}. Finally, we summarize
our findings in Sec.~\ref{s:results}.

%==================================== The Scalar Potential =====================
\section{The model}
\label{s:model}
The scalar sector of our model contains the following fields:
\begin{eqnarray}
%\label{}
\Phi = \left\{2,\frac{1}{2}\right\} \,; ~~ \chi = \{3,1\} \,; ~~ \kappa^{++} = \{1,2\} \,; ~~ \sigma = {\rm real ~ singlet} \,,
\end{eqnarray}
where, the numbers inside the curly brackets associated with the fields represent their 
transformations properties under $SU(2)_L$ and $U(1)_Y$ respectively.
The normalization for the hypercharge is such that the electric charges of the component
fields are given by, $Q=T_3+Y$. The fields, $\chi$ and $\sigma$ are odd under an additional $Z_2$ symmetry which has been introduced to prevent the occurrence of tree-level neutrino masses as well as to ensure the stability of the DM particle. The most general scalar potential involving these fields is given below:
\begin{eqnarray}
\label{e:potential}
V &=& -m_{\Phi}^2\left(\Phi^\dagger\Phi\right) +m_{\chi}^2\Tr\left(\chi^\dagger\chi\right) +m_{\kappa}^2|\kappa|^2 + \frac{m_{\sigma}^2}{2}\sigma^2 +\lambda_{\Phi}\left(\Phi^\dagger\Phi\right)^2 +\lambda_{\chi}\left\{\Tr\left(\chi^\dagger\chi\right)\right\}^2 
 \nonumber \\
&&+\lambda'_{\chi}\Tr\left[\left(\chi^\dagger\chi\right)^2\right]   +\lambda_{\kappa}|\kappa|^4 +\lambda_{\sigma}|\sigma|^4 +\lambda_{\Phi\chi}\left(\Phi^\dagger\Phi\right)\Tr\left(\chi^\dagger\chi\right) +\lambda'_{\Phi\chi}\left(\Phi^\dagger\chi \chi^\dagger\Phi \right)   \nonumber \\
&&+\lambda_{\Phi\kappa}\left(\Phi^\dagger\Phi\right)|\kappa|^2 +\lambda_{\Phi\sigma}\left(\Phi^\dagger\Phi\right)\sigma^2 +\lambda_{\kappa\chi}|\kappa|^2\Tr\left(\chi^\dagger\chi\right) +\lambda_{\sigma\chi}\sigma^2\Tr\left(\chi^\dagger\chi\right)  \nonumber\\
&&+\lambda_{\sigma\kappa}|\kappa|^2\sigma^2 +\left\{\mu_\kappa \kappa^{++} \Tr\left(\chi^\dagger\chi^\dagger\right) +\lambda_6\sigma\Phi^\dagger\chi\widetilde{\Phi} + {\rm h.c.} \right\}  \,,
\end{eqnarray}
where `$\Tr$' represents the trace over $2\times2$ matrices and $\widetilde{\Phi}=i\sigma_2\Phi^*$, with  $\sigma_2$ being the second Pauli matrix. We can take all the
parameters in the potential to be real without any loss of generality. 

For the leptonic Yukawa sector, we have the following Lagrangian:
\begin{eqnarray}
\label{e:yukawa}
{\mathscr L}_Y = - (\overline{L_L})_a (Y_e)_{ab} (\ell_R)_b \Phi + f_{ab} \ell_a^T C^{-1}(\ell_R)_b \kappa^{++} + {\rm h.c.}\,,
\end{eqnarray}
where, $L_L = (\nu_\ell,~ \ell)_L^T$ denotes the left-handed lepton doublet and $\ell_R$
represents the right-handed charged lepton singlet. $C$ is the charge conjugation
operator. We choose to work in the mass basis
of the charged leptons which means, $Y_e$ is a diagonal matrix with positive entries and
$f$ is a complex symmetric matrix with three unphysical phases.

\subsection{The scalar spectrum}
\begin{table}[htbp!]
%\scriptsize
\begin{center}
{\tabulinesep=1.2mm
\begin{tabu}{|c|c|}
\hline
$Z_2$-even particles & $Z_2$-odd particles  \\
\hline\hline
SM fermions and gauge bosons, $h$ and $\kappa^{\pm\pm}$        &  $S$, $A$, $H$, $\chi^\pm$, $\chi^{\pm\pm}$  \\
\hline
\end{tabu}
}
\end{center}
\caption{$Z_2$ parity assignments to the physical particles in our model.}
\label{t:z2}
\end{table}

We do not want to break the $Z_2$ symmetry spontaneously. Denoting by $v$ the vacuum expectation values (vev) of the doublet the minimization conditions read
\begin{eqnarray}
m_{\Phi}^2&=& \lambda_{\Phi}v^2  \,.
\end{eqnarray}
After spontaneous symmetry breaking~(SSB) we represent the doublet and the triplet as follows:
\begin{eqnarray}
%\label{}
\Phi = \frac{1}{\sqrt{2}} \begin{pmatrix} \sqrt{2} \omega^+ \\ v+h+i\zeta \end{pmatrix} \,, &&
\chi = \frac{1}{\sqrt{2}} \begin{pmatrix}  \chi^+ & \sqrt{2}\chi^{++} \\ h_t+iA & -\chi^+ \end{pmatrix} \,,
\end{eqnarray}
where, $\omega$ and $\zeta$ represent the Goldstones associated with the $W$ and
 $Z$ bosons respectively.
Because of the unbroken $Z_2$ symmetry, only $h_t$ and $\sigma$ can have nontrivial
mixing. This leads to a very simple scalar spectrum as described below.

%\emph{Doubly charged scalars:}
The masses for the doubly charged particles are given by,
\begin{eqnarray}
\mkpp^2 = m_\kappa^2 +\frac{1}{2}\lambda_{\Phi\kappa} v^2 \,, ~~
\mcpp^2 = m_\chi^2+\frac{1}{2}\lambda_{\Phi\chi}v^2 \,.
\label{m:doubly}
\end{eqnarray}
%
%\emph{Singly charged scalar:}
The mass of the singly charged scalar is given by,
\begin{eqnarray}
\mcp^2 = m_\chi^2+\frac{1}{4}(2\lambda_{\Phi\chi}+\lambda'_{\Phi\chi})v^2 \,.
\label{m:singly}
\end{eqnarray}
%
%\emph{CP odd scalar:}
The pseudoscalar mass is given by,
\begin{eqnarray}
m_A^2 = m_\chi^2+\frac{1}{2}(\lambda_{\Phi\chi}+\lambda'_{\Phi\chi})v^2 \,.
\label{m:pseudo}
\end{eqnarray}
From Eqs.~(\ref{m:doubly}), (\ref{m:singly}) and (\ref{m:pseudo}) it is easy to see that the following correlation holds:
\begin{eqnarray}
\label{e:corr}
m_{\chi+}^2-m_{\chi++}^2 = m_{A}^2-m_{\chi+}^2 = \frac{1}{4}\lambda'_{\Phi\chi}v^2 \,.
\end{eqnarray}

%\emph{CP even scalars:}
In the CP even sector, the SM-like Higgs arises purely from the doublet, $\Phi$, with mass
$m_h^2 = 2\lambda_\Phi v^2$. For the other two $Z_2$-odd scalars, we obtain the following
mass matrix:
\begin{eqnarray}
&& V_{\rm mass}^{\rm S} = \frac{1}{2} \begin{pmatrix}  \sigma & h_t \end{pmatrix} \begin{pmatrix}
A & -B   \\ -B & C  \end{pmatrix} \begin{pmatrix} \sigma \\ h_t \end{pmatrix}~~ {\rm with,} \\
&& A= m_\sigma^2+\lambda_{\Phi\sigma}v^2 \,, ~~
B= - \frac{1}{\sqrt{2}}\lambda_6v^2 \,, ~~
C=m_\chi^2+\frac{1}{2}(\lambda_{\Phi\chi}+\lambda'_{\Phi\chi})v^2 \,.
\label{e:ABC}
\end{eqnarray}
This mass matrix can be diagonalized by the following orthogonal rotation:
\begin{subequations}
%\label{}
\begin{eqnarray}
\begin{pmatrix} S \\ H \end{pmatrix} &=& \begin{pmatrix} \cos\alpha & -\sin\alpha \\ \sin\alpha & \cos\alpha \end{pmatrix} \begin{pmatrix}  \sigma \\ h_t \end{pmatrix} \,, \\
{\rm with,}~~ m^2_{H,S} &=& \frac{1}{2}\left\{(A+C)\pm\sqrt{(A-C)^2+4B^2}\right\} \,, \\
{\rm and,}~~ \tan2\alpha &=& \frac{2B}{A-C} \,,
%\label{}
\end{eqnarray}
\end{subequations}
where we have implicitly assumed that `$S$' is the lighter mass eigenstate. One can easily find the following relations:
\begin{subequations}
\label{e:abc}
\begin{eqnarray}
A&=& m_H^2\sin^2\alpha+m_S^2\cos^2\alpha \,, \\
C &=& m_H^2\cos^2\alpha +m_S^2\sin^2\alpha =  m_A^2 \,, \label{e:mA}\\
B &=& - \sin\alpha\cos\alpha (m_H^2-m_S^2) \,, \label{e:B}
\end{eqnarray}
\end{subequations}
which imply,
\begin{eqnarray}
\label{e:corr2}
m_S < m_A < m_H \,.
\end{eqnarray}
Combining \Eqs{e:ABC}{e:B} we can express $\lambda_6$ in terms of the physical parameter as follows:
\begin{eqnarray}
%\label{}
\lambda_6 = \frac{\sqrt{2}\sin\alpha\cos\alpha}{v^2}\left(m_H^2-m_S^2\right) \,.
\end{eqnarray}

The splittings between different scalar masses can be constrained further from the electroweak $T$-parameter.
The expression for the new physics contribution to the $T$-parameter is given by
\begin{eqnarray}
\label{e:T}
\Delta T &=& \frac{1}{4\pi\sin^{2}\theta_{W}M_{W}^{2}} \left[F(m_{\chi^{++}}^{2},m_{\chi^{+}}^{2})+\frac{1}{2} F(m_{\chi^{+}}^{2},m_{A}^{2})     \right. \nonumber \\
&& \left. + \frac{1}{2} \cos^{2}\alpha\left\{F(m_{\chi^{+}}^{2},m_{H}^{2})-2F(m_{A}^{2},m_{H}^{2})\right\} + \frac{1}{2}\sin^{2}\alpha\left\{F(m_{\chi^{+}}^{2},m_{S}^{2})-2F(m_{A}^{2},m_{S}^{2}) \right\} \right]\,,
\end{eqnarray}
where, $\theta_{W}$ and $M_{W}$ are the weak mixing angle and the
$W$-boson mass respectively. The function, $F(m_1^2,m_2^2)$, is given by,
\begin{eqnarray}
\label{e:fxy}
F(m_{1}^{2},m_{2}^{2})\equiv \frac{1}{2} 16\pi^{2}\int\frac{\dd[4]{k}}{(2\pi)^{4}}k^{2} \left(\frac{1}{k^{2}+m_{1}^{2}}-\frac{1}{k^{2}+m_{2}^{2}}\right)^{2}=\frac{m_{1}^{2}+m_{2}^{2}}{2}-\frac{m_{1}^{2}m_{2}^{2}}{m_{1}^{2}-m_{2}^{2}}\log\left(\frac{m_{1}^{2}}{m_{2}^{2}}\right)\,.
\end{eqnarray}
Taking the new physics contribution to the $T$-parameter as\cite{Baak:2013ppa}
\begin{eqnarray}
%\label{}
\Delta T = 0.05 \pm 0.12  \,,
\end{eqnarray}
we will require our model value of the $T$-parameter to be within the $2\sigma$ uncertainty range. For 
small $\sin\alpha$, this leads to  $|m_H-\mcpp|\lesssim 100$~GeV.

In passing, combining \Eqs{e:corr}{e:corr2}, we note that two types of scalar mass hierarchies are possible
depending on the sign of $\lambda'_{\Phi\chi}$,
\begin{subequations}
\label{e:hierarchies}
\begin{eqnarray}
&& m_H > m_A > \mcp > \mcpp > m_S \,, \\
{\rm or,} && \mcpp > \mcp > m_A > m_S ~~ {\rm and} ~~ m_H > m_A \,.
\end{eqnarray}
\end{subequations}
In both cases, $\mkpp$ can be arbitrary in principle.

\section{Estimation of \texorpdfstring{$0\nu\beta\beta$}{TEXT} decay}
\label{s:nbb}
\begin{figure}%[htbp!]
\begin{centering}
\includegraphics[scale=0.55]{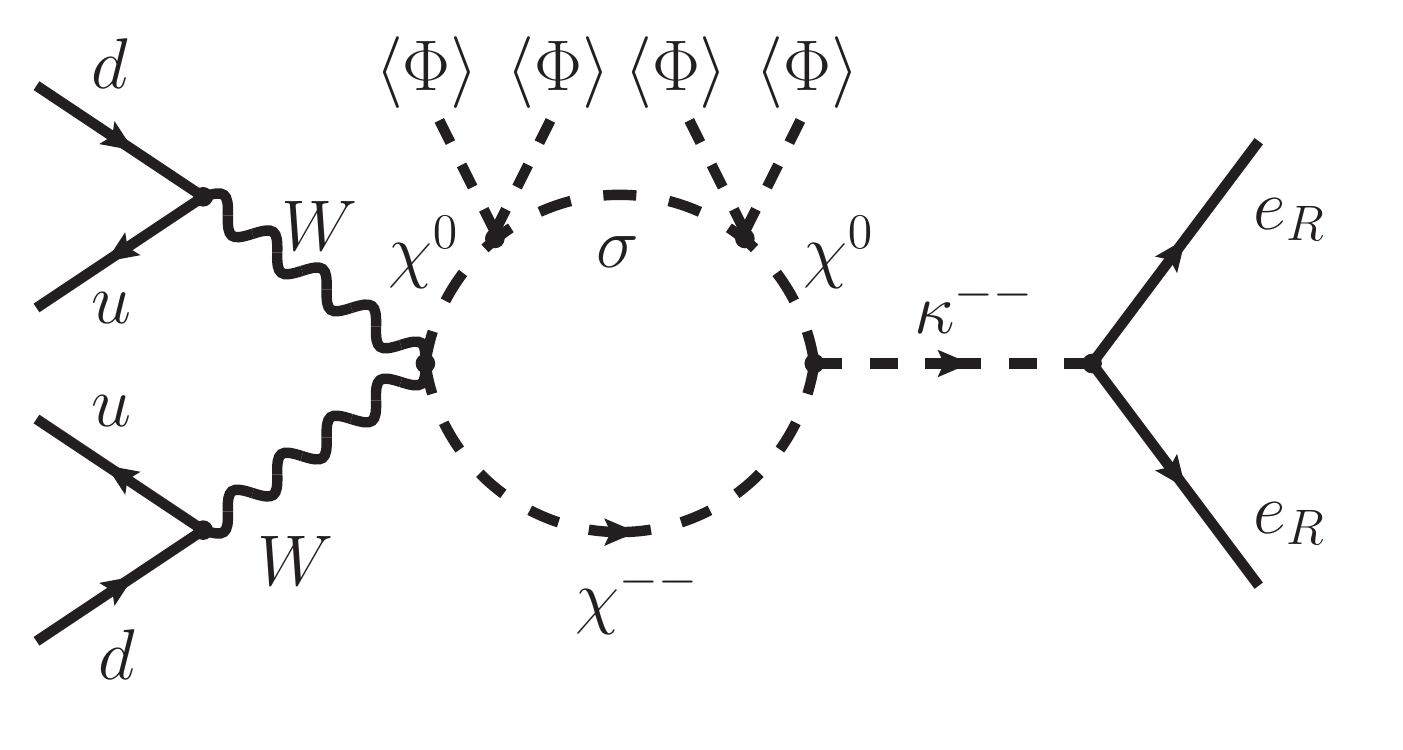}
\par
\end{centering}
\caption{One-loop diagram, in the mass insertion approach, contributing to
neutrinoless double beta decay.\label{fig:0nu2beta}}
\end{figure}
For new scalar masses of $\order(1~{\rm TeV})$, the Majorana mass matrix element,
$M_{ee}$, will be very small (see Sec.~\ref{sec:nu-mass} for details).
As a result, the usual neutrino exchange diagram will contribute negligibly
to $\nbb$. The main contribution to the $\nbb$ amplitude has been displayed in Fig.~\ref{fig:0nu2beta}.
From the diagram in Fig.~\ref{fig:0nu2beta} we can easily estimate
the effective $\bar{e}e^{c}(\bar{u}d)^{2}$ interaction giving rise
to $\nbb$ 
\begin{equation}
\mathcal{L}_{0\nu\beta\beta}=2\frac{f_{ee}^{*}}{16\pi^{2}}\frac{\mu_{\kappa}\lambda_{6}^{2}}{\mkpp^{2}m_{A}^{4}}I_{\beta}\left(\overline{u_{L}}\gamma^{\mu}d_{L}\right)\left(\overline{u_{L}}\gamma_{\mu}d_{L}\right)\overline{e_{R}}e_{R}^{c} \,,
\label{eq:Lagrangian0n2bModel}
\end{equation}
where $I_{\beta}$ is a dimensionless function of the scalar masses
running in the loop which is expected to be $\order(1)$. For illustration, we have
chosen the common scale of the loop to be the mass of the pseudoscalar
part from the scalar triplet, $m_{A}$. Of
course the diagram in Fig.~\ref{fig:0nu2beta} is only one of the
contributions in the mass insertion approach which allows us to give
an estimate. A complete calculation of the function $I_\beta$  in the
physical basis has been presented in Appendix~\ref{ap:Neutrinoless} yielding values for $I_\beta$ which are slightly smaller than one in the range of masses of interest, $I_\beta\sim 0.1$. We will use these values for our estimates.

The interaction of \Eqn{eq:Lagrangian0n2bModel} has been considered in the literature~\cite{Pas:2000vn,Deppisch:2012nb},
where it was parametrized as follows:
\begin{equation}
\mathcal{L}_{0\nu\beta\beta}=\frac{G_{F}^{2}}{2m_{p}}\epsilon_{3}\left(\bar{u}\gamma^{\mu}(1-\gamma_{5})d\right)\left(\bar{u}\gamma_{\mu}(1-\gamma_{5})d\right)\bar{e}(1-\gamma_{5})e^{c}\,.
\label{eq:Lagrangian-0nu2beta}
\end{equation}
Comparing \Eqs{eq:Lagrangian0n2bModel}{eq:Lagrangian-0nu2beta} we obtain,
\begin{equation}
\epsilon_{3}=\frac{m_{p}}{2G_{F}^{2}}\frac{f_{ee}^{*}}{16\pi^{2}}\frac{\mu_{\kappa}\lambda_{6}^{2}}{\mkpp^{2}m_{A}^{4}}I_{\beta}\,.
\label{eq:epsilon3}
\end{equation}

In Ref.~\cite{Deppisch:2012nb}, to set bounds on $\epsilon_3$, the authors used the limits on the
half-life for the $\nbb$ from the most sensitive experiments of that time, namely,
$T_{1/2}^{0\nu\beta\beta}(^{76}\mathrm{Ge})>1.9\times10^{25}$~yrs
(HM \cite{KlapdorKleingrothaus:2000sn}) and $T_{1/2}^{0\nu\beta\beta}(^{136}\mathrm{Xe})>1.6\times10^{25}$~yrs
(EXO-200 \cite{Auger:2012ar}). However KamLAND-Zen has recently obtained a stronger limit on
the lifetime from $^{136}\mathrm{Xe}$, $T_{1/2}^{0\nu\beta\beta}(^{136}\mathrm{Xe})>1.07\times10^{26}\,\mathrm{yr}$
\cite{KamLAND-Zen:2016pfg}, which, using the matrix elements from
\cite{Deppisch:2012nb}, translates to $\epsilon_{3}<4\times10^{-9}$ at 90\%~C.L.

On the other hand, upcoming experiments are expected to be sensitive
to lifetimes of order $10^{27}$--$10^{28}$~yrs\cite{Barabash:2011fg},
\emph{i.e.} a reduction factor on the coupling of about one order
of magnitude. Thus, for $\nbb$ mediated by
heavy particles to be observable in the next round of experiments we
should have $\epsilon_{3}\gtrsim4\times10^{-10}$. Therefore in order to escape the current
experimental bounds but at the same time to entertain the possibility of observing $\nbb$
in the near future, we require $\epsilon_3$ to be within the following range:
\begin{eqnarray}
\label{e:epsrange}
4\times 10^{-10} < \epsilon_3 < 4\times10^{-9} \,.
\end{eqnarray}

With $f_{ee},~\lambda_{6}\approx\text{1}$, $\mu_{\kappa}\approx m_{A}\approx \mkpp\approx1\,\mathrm{TeV}$
and $I_{\beta}\sim 0.1$ we obtain, from \Eqn{eq:epsilon3}, $\epsilon_{3}\sim10^{-9}$ which falls
naturally within the range given in \Eqn{e:epsrange}.

%%%%%%%%%%%%%%%%%%%%%%%%%%%%%%%%%%%%%%%%%%%%%%%%%%%%%%%%%%%%%%%
%%%%%%%                    Section: Neutrino masses                  %%%%%%%%%%%%%%%%%%%%%%%%%%%%%
%%%%%%%%%%%%%%%%%%%%%%%%%%%%%%%%%%%%%%%%%%%%%%%%%%%%%%%%%%%%%%%
\section{Estimation of the neutrino masses}
\label{sec:nu-mass}
\begin{figure}[htbp!]
\centering
\includegraphics[scale=0.55]{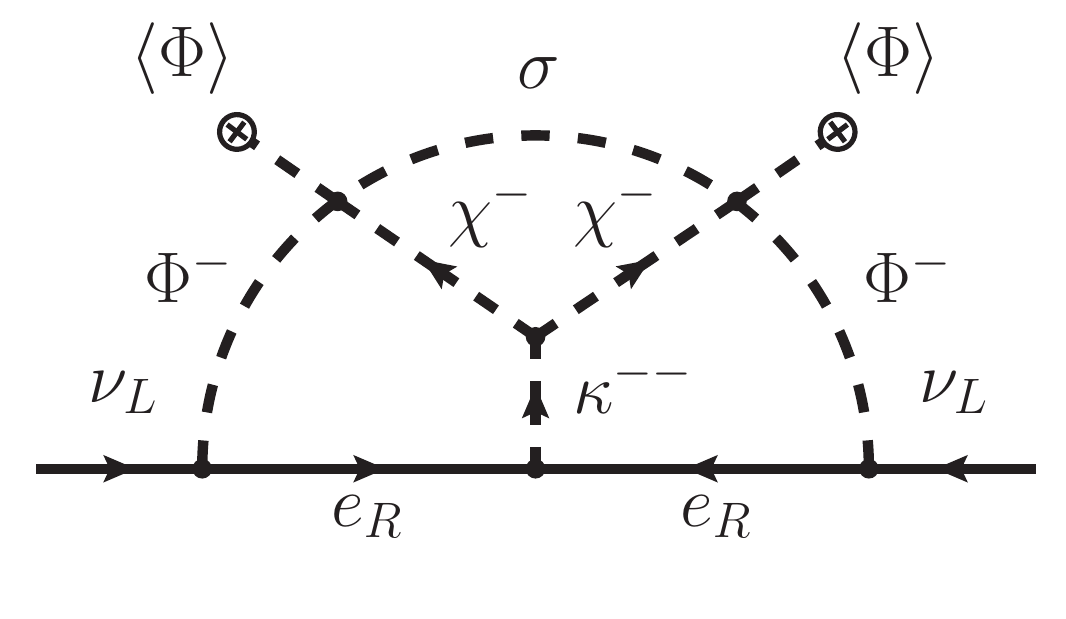}
\caption{Sample three loop diagram, in the mass insertion approach, contributing to the neutrino masses.}
\label{fig:numass}
\end{figure}

From \Eqs{e:potential}{e:yukawa} it is obvious that simultaneous nonzero values for 
$Y_e$, $f_{ab}$, $\mu_\kappa$ and $\lambda_6$ will prevent us from assigning consistent lepton numbers to all the scalar
and lepton fields. Therefore, lepton number is broken explicitly and Majorana neutrino masses will be unavoidable. The sample diagram of Fig.~\ref{fig:numass}, in the mass insertion approach, clearly
depicts the involvement of all these couplings in a multiplicative manner. Thus, we can parametrize
the neutrino mass matrix as follows:
\begin{eqnarray}
\label{e:nuelements}
M_{ab}=\frac{8\mu_{\kappa}\lambda_{6}^{2}}{(4\pi)^{6}\mkpp^{2}}I_{\nu}\, m_{a}f_{ab}m_{b} \,,
\end{eqnarray}
where $m_{a}$ denotes the mass of the charged lepton, $\ell_a$, and $I_{\nu}$
represents the loop function expected to be of $\order(1)$. Detailed expression of $I_\nu$
in terms of the scalar masses has been presented in Appendix~\ref{ap:Neutrino-masses}. Eq.~(\ref{e:nuelements}) has a very particular and predictive structure, specific for this class of models, which can be constrasted with the observed spectrum of neutrino masses and mixings (see for instance Refs. \cite{delAguila:2011gr,delAguila:2012nu,Gustafsson:2014vpa}).  

As before, taking $f_{\tau\tau},~\lambda_{6}\approx 1$ and $\mu_{\kappa}\approx \mkpp\approx1$~TeV
and $I_{\nu}\sim 1$ we obtain the following values for the different elements
\begin{eqnarray}
%\label{}
M_{ee}\sim10^{-7}\,\mathrm{eV}\,, ~~ M_{e\mu}\sim10^{-4}\,\mathrm{eV}\,, ~~ 
M_{e\tau}\sim10^{-3}\,\mathrm{eV}\,,~~ M_{\mu\mu}\sim10^{-2}\,\mathrm{eV}\,, ~~ M_{\mu\tau}\sim10^{-1}\,\mathrm{eV}\,,~~
M_{\tau\tau}\sim 10\,\mathrm{eV}\,.
\end{eqnarray}
But of course, some of the $f_{ab}$s can be much smaller than~1. However, not all of the elements of the $f$ matrix are
arbitrary as some of them will be constrained from LFV processes. We will discuss these constraints
in Sec.~\ref{s:LFV}. But for now we wish to emphasize that the product $|f_{ee}^*f_{e\mu}|$ will receive strong
bounds from $\mu\to 3e$ as the latter can proceed at the tree-level mediated by $\kappa^{++}$. Then, one should
naturally expect the following hierarchy among the mass matrix elements:
\begin{eqnarray}
\label{e:hi1}
M_{ee},M_{e\mu}\ll M_{e\tau},M_{\mu\mu},M_{\mu\tau},M_{\tau\tau} \,,
\end{eqnarray}
which, obviously, can only accommodate a normal hierarchy among the neutrino masses. In Ref.~\cite{delAguila:2011gr} it has been shown that the above hierarchy with
\begin{eqnarray}
\label{e:hi2}
3 M_{e\tau} \sim M_{\mu\mu} \sim M_{\mu\tau} \sim M_{\tau\tau} \sim 0.02~{\rm eV} 
\end{eqnarray}
can successfully reproduce the observed masses and mixings in the neutrino sector with
a prediction of $\sin^2\theta_{13} > 0.008$. \Eqn{e:hi2} will imply the following
hierarchy among the Yukawa elements:
\begin{eqnarray}
\label{e:hi3}
3 f_{e\tau} \sim \frac{m_\tau}{m_e}f_{\tau\tau} > f_{\mu\mu} \sim \frac{m_\tau^2}{m_\mu^2}f_{\tau\tau}
> f_{\mu\tau} \sim \frac{m_\tau}{m_\mu}f_{\tau\tau} > f_{\tau\tau} \,.
\end{eqnarray}
We shall also assume $f_{ee}\gg f_{e\mu}$ in such a way that $f_{ee}^*f_{e\mu}$ is still sufficiently small
to keep $\mu\to 3e$ decay under control but at the same time allowing for the possibility of large $\nbb$.

From \Eqs{eq:epsilon3}{e:nuelements} we see that the dimensionless factor,
\begin{eqnarray}
\label{e:gamma}
\gamma = \frac{\mu_\kappa \lambda_6^2}{\mkpp} = \frac{2\sin^2\alpha\cos^2\alpha(m_H^2-m_S^2)^2}{v^4} \frac{\mu_\kappa}{\mkpp} \,,
\end{eqnarray}
is common to both. In terms of $\gamma$, the explicit expression for $M_{\tau\tau}$ in \Eqn{e:hi2} reads:
\begin{eqnarray}
\label{e:mtt}
M_{\tau\tau} = \frac{8}{(4\pi)^6} \gamma I_\nu \frac{m_\tau^2f_{\tau\tau}}{\mkpp} \approx 0.02 ~{\rm eV} \,.
\end{eqnarray}
As we will see in Sec.~\ref{s:LFV}, the ratio $f_{\tau\tau}/\mkpp$ is bounded from LFV processes as
$f_{\tau\tau}/\mkpp\lesssim 1.4\times 10^{-4}~{\rm TeV}^{-1}$. Plugging this into \Eqn{e:mtt} we obtain the following
bound for $\gamma$:
\begin{eqnarray}
\label{e:gambound}
\gamma \gtrsim \frac{22}{I_\nu} \,.
\end{eqnarray}

Having an explicit expression for the neutrino masses we can compare the light neutrino exchange contributions to $\nbb$ with the ones discussed in Sec.~\ref{s:nbb}.
In fact, from \Eqs{e:nuelements}{eq:epsilon3} we can express the neutrino mass matrix element $M_{ee}$, which controls the $\nu$ contributions to $\nbb$, in terms $\epsilon_3$, which parametrizes the new contributions
\begin{equation}\label{e:Mee-epsilon3}
M_{ee}=\frac{16m_{e}^{2}G_{F}^{2}m_{A}^{4}}{m_{p}(4\pi)^{4}}\frac{I_{\nu}}{I_{\beta}}\epsilon_{3}~.
\end{equation}
Then, it is clear that for small enough $m_A$ the new contributions will dominate over the neutrino contributions. How small? Since the nuclear matrix elements are different in the two cases we cannot make a direct comparison. However, we can use that the experimental limit 
$T_{1/2}^{0\nu\beta\beta}(^{136}\mathrm{Xe})>1.07\times10^{26}\,\mathrm{yrs}$\cite{KamLAND-Zen:2016pfg}
translates into two equivalent bounds on $\epsilon_3$ and $M_{ee}$ when $\nbb$ is dominated by the new contributions or by neutrino masses respectively:
     \begin{eqnarray}
     \label{e:f1}
     \epsilon_3 < 4\times 10^{-9} \,, ~~ M_{ee} < 0.1~{\rm eV} \, ,
     \end{eqnarray}
which already include the appropriate nuclear matrix elements. Using these results and taking $I_\beta \sim 0.1 I_\nu$ we obtain that the new contributions will dominate for $m_A \lesssim 15$~TeV. Therefore,  scalar masses must be relatively light, and this could make the model testable at the LHC and/or in LFV processes. 
% Denoting the hadronic matrix element corresponding to the operator in \Eqn{eq:Lagrangian-0nu2beta} by $x_3$
% and that corresponding to the conventional $\nu_L$ mediated diagram by $x_{ee}$, we may write:
% \begin{eqnarray}
%      \label{e:f2}
%      x_3 \left|\epsilon_3^{\rm max}\frac{G_F^2}{2m_p^2 }\right|^2 = x_{ee} |M_{ee}^{\rm max}|^2 \,.
%      \end{eqnarray}
% For the diagram in Fig.~\ref{fig:0nu2beta} to dominate we will need
%      \begin{eqnarray}
%      \label{e:f3}
%      x_3 \left|\epsilon_3\frac{G_F^2}{2m_p^2 }\right|^2 > x_{ee} |M_{ee}|^2 \,.
%      \end{eqnarray}
% The above relation in conjunction with Eqs.~(\ref{eq:epsilon3}), (\ref{e:nuelements}) and (\ref{e:f2}) will lead to the bound in \Eqn{e:malimit}.
% 
% %
% \begin{eqnarray}
% \label{e:malimit}
% m_{A}<4\pi\left[ \left(\frac{0.1~{\rm eV}}{4\times 10^{-9}}\right) \frac{I_{\beta}}{I_{\nu}}\frac{m_{p}}{16m_{e}^{2}G_{F}^{2}}\right]^{1/4}\sim30\,\mathrm{TeV} \,,
% \end{eqnarray}
% where, for this naive estimate, we have taken $I_{\beta}/I_{\nu}\sim1$. 
%This suggests that, in this model, for $\nbb$ to be
%observable in the next round of experiments we would need the new scalars to be relatively light.

% % % % % % % % % % % % % % % % % % % % % % % % % % % % % % % % % % % % % % % % % % % % % % % % % %
% % % % % % %   Constraints from LFV processes  % % % % % % % % % % % % % % % % % % % % % % % % % % % % %
% % % % % % % % % % % % % % % % % % % % % % % % % % % % % % % % % % % % % % % % % % % % % % % % % %
%
\begin{table}[htbp!]
%\scriptsize
\begin{center}
{\tabulinesep=1.2mm
\begin{tabu}{|c|c|c|}
\hline
%\multicolumn{1}{|c|}{Experimental Data (90\% CL)} 
Experimental Data (90\% CL) & Bounds (90\% CL) & Bounds assuming \Eqn{e:hi3} \\
\hline\hline
$\mathrm{BR}(\mu^{-}\rightarrow e^{+}e^{-}e^{-})<1.0\times10^{-12}$ & $|f_{e\mu}f_{ee}^{*}|<2.3\times10^{-5}\,\left(\frac{\mkpp}{\mathrm{TeV}}\right)^{2}$ & \\
\hline
$\mathrm{BR}(\tau^{-}\rightarrow e^{+}e^{-}e^{-})<2.7\times10^{-8}$ & $|f_{e\tau}f_{ee}^{*}|<0.009\,\left(\frac{\mkpp}{\mathrm{TeV}}\right)^{2}$ & $|f_{ee}^*f_{\tau\tau}|\lesssim 7.8\times10^{-6}\,\left(\frac{\mkpp}{\mathrm{TeV}}\right)^{2}$ \\
\hline
$\mathrm{BR}(\tau^{-}\rightarrow e^{+}e^{-}\mu^{-})<1.8\times10^{-8}$ & $|f_{e\tau}f_{e\mu}^{*}|<0.005\,\left(\frac{\mkpp}{\mathrm{TeV}}\right)^{2}$ & $|f_{e\mu}^*f_{\tau\tau}|\lesssim 4.3\times10^{-6}\,\left(\frac{\mkpp}{\mathrm{TeV}}\right)^{2}$ \\
\hline
$\mathrm{BR}(\tau^{-}\rightarrow e^{+}\mu^{-}\mu^{-})<1.7\times10^{-8}$ & $|f_{e\tau}f_{\mu\mu}^{*}|<0.007\,\left(\frac{\mkpp}{\mathrm{TeV}}\right)^{2}$ & $|f_{\tau\tau}|\lesssim 1.4\times10^{-4}\,\left(\frac{\mkpp}{\mathrm{TeV}}\right)$ \\
\hline
$\mathrm{BR}(\mu\rightarrow e\gamma)<5.7\times10^{-13}$ & \specialcell{$|f_{ee}^{*}f_{e\mu}+f_{e\mu}^{*}f_{\mu\mu}+f_{e\tau}^{*}f_{\mu\tau}|^{2}$ \\ $<1 \times10^{-7}\,(\frac{\mkpp}{\mathrm{TeV}})^{4}$} & $|f_{\tau\tau}|\lesssim 1.2\times10^{-4}\,\left(\frac{\mkpp}{\mathrm{TeV}}\right)$ \\
\hline
\end{tabu}
}
\end{center}
\caption{Relevant constraints for our model from LFV
decays~\cite{Olive:2016xmw,Adam:2013mnn}. Limits on the Yukawa couplings of the doubly charged
singlet scalars have been taken from Ref.~\cite{Herrero-Garcia:2014hfa}. The constraints in the third
column are obtained from those in the second column assuming \Eqn{e:hi3} holds. The bound in
the third column corresponding to $\mu\to e\gamma$ has an additional assumption, $f_{e\mu}\approx 0$.}
\label{tab:LFV}
\end{table}
\section{Constraints from LFV processes}
\label{s:LFV}

Constraints from LFV processes come mainly from decays of the type $\ell_a^\mp \to \ell_b^\pm\ell_c^\mp\ell_d^\mp$ and
$\ell_a^\mp \to \ell_b^\mp \gamma$. In our case $\ell_a^\mp \to \ell_b^\pm\ell_c^\mp\ell_d^\mp$ will be more important
because these decays can occur at the tree-level through the exchange of the doubly charged scalar singlet, $\kappa^{\pm\pm}$.
These processes along with the kinds of constraints they imply have been reviewed in Ref.~\cite{Herrero-Garcia:2014hfa} in the context of the Zee-Babu model (see also Refs. \cite{Babu:2002uu,Nebot:2007bc}).
The experimental data has not changed much since then. In the first two columns of Table~\ref{tab:LFV} we have summarized
the experimental data and the corresponding constraints on the Yukawa couplings. In the third column of Table~\ref{tab:LFV}
we recast the constraints of the second column assuming the validity of \Eqn{e:hi3}. This allows us to express the constraints
in more specific forms. For example, using $m_ef_{e\tau} \sim m_\tau f_{\tau\tau}$ and $m_\mu^2f_{\mu \mu} \sim 
m_\tau^2f_{\tau\tau}$, the constraint from $\tau\to e\mu\mu$ leads to a direct bound on $f_{\tau\tau}$ as follows:
\begin{eqnarray}
\label{e:LFV1}
|f_{\tau\tau}|\lesssim 1.4\times 10^{-4}\,\left(\frac{\mkpp}{\mathrm{TeV}}\right) \,.
\end{eqnarray}
It is also worth mentioning that, using \Eqn{e:hi3}, the limit from $\tau \to 3e$ translates into
\begin{eqnarray}
\label{e:LFV2}
|f_{ee}^*f_{\tau\tau}|\lesssim 7.8 \times 10^{-6}\,\left(\frac{\mkpp}{\mathrm{TeV}}\right)^{2} \,.
\end{eqnarray}
As mentioned earlier, we want to have $f_{ee}$ relatively large to have appreciable $\nbb$ rate in the future experiments.
Then we will need $f_{e\mu}$ to be vanishingly small to keep the constraints from $\mu\to 3e$ under control. Note
that, for $f_{ee}\sim \order(1)$ and sub TeV $\kappa^{++}$, \Eqn{e:LFV2} will imply a stronger bound on $f_{\tau\tau}$
than \Eqn{e:LFV1}.

% % % % % % % % % % % % % % % % % % % % % % % % % % % % % % % % % % % % % % % % % % %
% % % % % % % %     Dark Matter         % % % % % % % % % % % % % % % % % % % % % % % % % % %
% % % % % % % % % % % % % % % % % % % % % % % % % % % % % % % % % % % % % % % % % % % %
\section{Dark Matter }
\label{s:DM}
\begin{figure}[htbp!]
\centering
\includegraphics[scale=0.35]{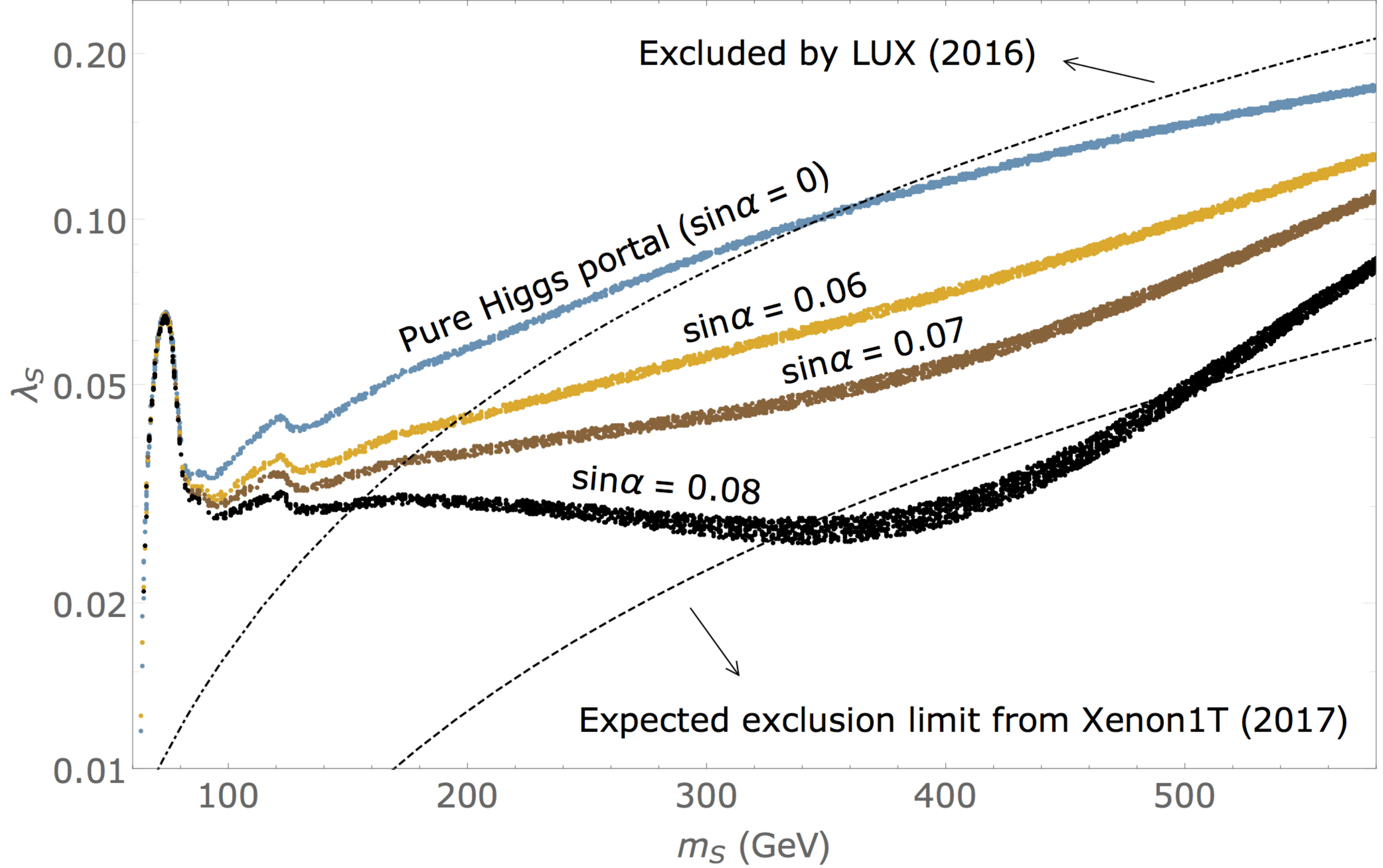}
\caption{Regions corresponding to the observed relic abundance\cite{Ade:2013zuv} in the $m_S$-$\lambda_S$ plane for different values of $\sin\alpha$. We have chosen $m_H=\mcpp=\mkpp=800$~GeV as a benchmark for this plot.
Current\cite{Tan:2016zwf,Akerib:2016vxi} and future\cite{Aprile:2015uzo} bounds
from direct detection experiments are also marked appropriately.}
\label{f:DM}
\end{figure}
Our model has a $Z_2$ symmetry which remains unbroken after the SSB. Consequently, the particle spectrum can be
divided into $Z_2$-even and odd sectors as shown in Table~\ref{t:z2}. Among the $Z_2$ odd neutral scalars, $S$, being
the lightest, is a promising candidate for DM. Notice that $S$ is and admixture of the real singlet and the triplet, and therefore, it will feel both, Higgs and gauge interactions\footnote{For recent studies of a DM candidate which is an admixture of a scalar singlet and a Y=0 triplet see for instance \cite{Fischer:2013hwa,Cheung:2013dua}.}. In spite of that, one can parametrize its couplings with the SM-like Higgs boson as follows:
\begin{eqnarray}
\label{e:lslag}
&& \ml \supset -\frac{1}{2} \lambda_{S}S^{2}\abs{\Phi^{0}}^{2} \supset 
-\frac{1}{2}\lambda_{S}S^{2}\left(vh+\frac{1}{2} h^{2}\right)\,, \\
\label{e:ls}
{\rm with,} && \lambda_{S} =  \frac{1}{2}\left[2\lambda_{\Phi\sigma} \cos^{2}\alpha -2\sqrt{2}\lambda_{6}\sin\alpha\cos\alpha +(\lambda_{\Phi\chi}+\lambda_{\Phi\chi}^{\prime}) \sin^{2}\alpha\right]\,.
\end{eqnarray}

In Fig.~\ref{f:DM} we have displayed regions in the $m_S$-$\lambda_S$ plane, which can reproduce the observed DM
relic density\cite{Ade:2013zuv}. For this plot, we have assumed $m_H=\mcpp=\mkpp=800$~GeV and used the MicrOMEGAs
package\cite{Belanger:2013oya} to compute the DM abundance. Note that, the region labeled as $\sin\alpha=0$
corresponds to the pure Higgs portal scenario. Barring the small window near the Higgs-pole ($m_S\approx m_h/2$, not shown explicitly in the plot), in this case, we need $m_S\gtrsim 350$~GeV\cite{Han:2015hda,Escudero:2016gzx} to evade the direct search bound.
It is worth mentioning that in the case of pure Higgs portal, for our choice of benchmark, the DM annihilates through
$f\bar{f}$, $WW$, $ZZ$ and $hh$ mainly. All these annihilation channels except $hh$ can only proceed through
s-channel $h$ exchange. But as $\sin\alpha$ is turned on, we allow for a direct $SSVV$ ($V=W,Z$) with strength
proportional to $g^2\sin^2\alpha$. For our choice of positive values for $\lambda_S$, the new contact diagram will
interfere constructively with the $h$ mediated s-channel diagram.\footnote{
A nonzero value of $\sin\alpha$ will also induce t-channel diagrams for $SS\to VV,hh$ mediated by $\chi^\pm$, $A$ or $H$.
But these amplitudes will be suppressed as long as $\mcp,m_A, m_H\gg m_S$. Also note that, in this limit, the gauge couplings of $S$ do not contribute to the direct detection cross section\cite{TuckerSmith:2004jv,Cirelli:2005uq}.}
This will enhance the annihilation rate for $SS\to VV$ once the corresponding threshold is reached. Therefore, we would require lower values of $\lambda_S$, compared to the pure Higgs portal case, to reproduce the relic abundance. These features have been depicted in Fig.~\ref{f:DM} where we can see that a small value of $\sin\alpha$ is sufficient to accommodate DM with mass as low as $200$~GeV, which can either be discovered or ruled out in the next run of direct detection experiments.

% % % % % % % % % % % % % % % % % % % % % % % % % % % % % % % % % % % % % % % % % % % % % % %
% % % % % % % % % % % %     Conclusion      % % % % % % % % % % % % % % % % % % % % % % % % % % % %
% % % % % % % % % % % % % % % % % % % % % % % % % % % % % % % % % % % % % % % % % % % % % % %
\section{Results and conclusions}
\label{s:results}
\begin{table}%[htbp!]
%\scriptsize
\begin{center}
{\tabulinesep=1.2mm
\begin{tabu}{|c|c|c|c|c|c|c|c|c|}
\hline
 $\mcpp$ (GeV) & $\mkpp$ (GeV) & $\sin\alpha$ & $m_H$ (GeV) & $m_S$ (GeV) & $\mu_\kappa$ (TeV) & $\abs{f_{ee}}$ & $\abs{f_{\tau\tau}}$ & $\abs{f_{e\mu}}$ \\
\hline\hline
 800 & 800 & 0.08 & 800 & 200 & 20 & 0.01 & $10^{-4}$ & 0 \\
\hline
\end{tabu}

\vspace*{2mm}

\begin{tabu}{|c|c|c|c|c|c|c|c|}
\hline
 $\mcp$ (GeV) & $m_A$ (GeV) & $I_\beta$ & $I_\nu$ & $\epsilon_3$ & $\abs{f_{e\tau}}$  &  $\abs{f_{\mu\mu}}$ & $\abs{f_{\mu\tau}}$ \\
\hline\hline
799 & 798 & 0.165 & 0.84 & $3.5\times 10^{-9}$ & $0.12$ & $0.03$ & $1.7\times 10^{-3}$ \\
\hline
\end{tabu}
}
\end{center}
\caption{Benchmark values for the input parameters (first row) and other relevant quantities derived from these inputs (second row).}
\label{tab:inputoutput}
\end{table}
%

% \begin{table}[htbp!]
% %\scriptsize
% \begin{center}
% {\tabulinesep=1.2mm
% \begin{tabu}{|c|c|c|c|c|c|c|c|}
% \hline
%  $\mcp$ (GeV) & $m_A$ (GeV) & $I_\beta$ & $I_\nu$ & $\epsilon_3$ & $\abs{f_{e\tau}}$  &  $\abs{f_{\mu\mu}}$ & $\abs{f_{\mu\tau}}$ \\
% \hline\hline
% %
% 799 & 798 & 0.165 & 0.84 & $3.5\times 10^{-9}$ & $0.12$ & $0.03$ & $1.7\times 10^{-3}$ \\
% \hline
% \end{tabu}
% }
% \end{center}
% \caption{Other relevant parameters computed using the inputs from Table~\ref{tab:input}.}
% \label{tab:output}
% \end{table}
%
Since $\kappa^{\pm\pm}$ couples directly to the charged leptons, it will be strongly constrained from the same
sign dilepton searches at the LHC. Depending on the preferred decay channel of $\kappa^{\pm\pm}$, the bound
can be as strong as $\mkpp\gtrsim 500$~GeV\cite{CMS:2016cpz,ATLAS:2016pbt}. On the other hand, to keep
the $T$-parameter under control, for small $\sin\alpha$, we will need $|m_H-\mcpp|\lesssim 100$~GeV (see \Eqn{e:T}).
All these considerations together justify our choice of benchmark for Fig.~\ref{f:DM}. Now, to satisfy \Eqn{e:gambound}
we need to have a large splitting between $m_H$ and $m_S$. Keeping these things in mind, we have chosen the first row in
Table~\ref{tab:inputoutput} as a benchmark for the input parameters. Some relevant output quantities that follow from
these inputs have also been displayed in the second row of the same table. From the numbers of Table~\ref{tab:inputoutput}
one can easily check that the constraints of \Eqs{e:epsrange}{e:gambound} and all the
bounds in Table~\ref{tab:LFV} are satisfied. Moreover, using \Eqn{e:hi3} suitable values for $f_{e\tau}$,
$f_{\mu\mu}$ and $f_{\mu\tau}$ can be found so that the hierarchy of \Eqn{e:hi2} is satisfied.

\begin{figure}[htbp!]
\centering
\includegraphics[scale=0.70]{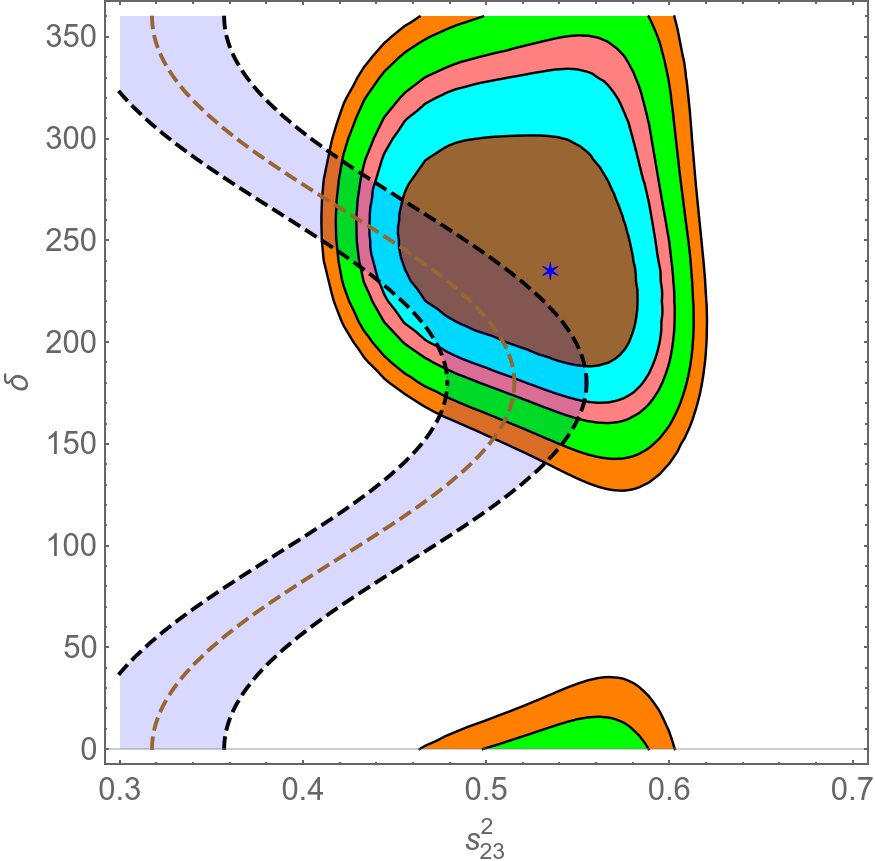}
\caption{The NuFIT results \cite{Esteban:2016qun,nufitweb18} for the global fit to neutrino data (coloured contours correspond to 68.27\%  90\%  95.45\%  99\%  99.73\% C.L. regions in the $s^2_{23}$--$\delta$ plane) against the prediction of the model for central values of the rest of the mixing parameters (brown dashed line) and the band obtained when they are varied in 1$\sigma$.}
\label{f:deltas23}
\end{figure}

The model has many phenomenological implications that make it special and distinguishable from similar models.
To exemplify one such feature, we note that the requirement,
 $M_{ee},M_{e\mu}\ll M_{e\tau},M_{\mu\mu},M_{\mu\tau},M_{\tau\tau}$, and consequently NH among the neutrino masses, results in a strong correlation between $\delta$, the CP violating phase of the Pontecorvo–Maki–Nakagawa–Sakata mixing matrix, and the other mixing parameters. For instance, in Fig. \ref{f:deltas23} we have displayed the allowed region in the plane $s^2_{23}$--$\delta$ obtained by the NuFIT collaboration (version 3.2 of 2018)\cite{Esteban:2016qun,nufitweb18} (the different coloured contours are 68.27\%,  90\%,  95.45\%,  99\% and  99.73\% C.L. regions respectively). On top of it we superimpose the correlation obtained from the requirement $M_{ee}=M_{e\mu}=0$ for the central values of the rest of the mixing parameters (brown dashed line) and the band obtained when they are varied in 1$\sigma$. 
% As we can see, the prediction of the model agrees very well with the central value of the fit. Therefore, adding the constraint from the model to the fit will
% strengthen the trend, already present in the data, towards values around $\delta = 260^\circ$ and $s^2_{23}$ in the first octant. 
% Moreover the model also predicts the smallest neutrino mass to be around $m_1 \sim 5\times 10^{-3}$~eV and the two Majorana phases $\alpha_1\sim 360^\circ-\delta\sim 100^\circ$ and $\alpha_2\sim \alpha_1+180^\circ \sim 280^\circ$
As we can see, the prediction of the model agrees well with the fit, although with some trend to lower values of $s^2_{23}$ and $\delta$. 
Moreover the model also predicts the smallest neutrino mass to be around $m_1 \sim 5\times 10^{-3}$~eV and the two Majorana phases $\alpha_1\sim 360^\circ-\delta\sim 130^\circ$ and $\alpha_2\sim \alpha_1+180^\circ \sim 310^\circ$
\footnote{Here we use the same conventions for the neutrino mixing phases used in Ref.~\cite{delAguila:2011gr} except that now we take them in the range $[0^\circ,360^\circ]$ in order to compare with NuFIT results.} 

\Eqn{e:nuelements} allows us to write the couplings $f_{ab}$ in terms of the neutrino masses and mixings up to a global factor. Since these couplings control all the LFV decays mediated by the double charged scalars, all the LFV processes are, in principle, predicted in terms of neutrino masses and mixing parameters which are fixed in our model.

As can be seen from the value of $\epsilon_3$ in Table~\ref{tab:inputoutput}, our model opens up the interesting possibility of detecting $\nbb$ in the next
generation of experiments even if $M_{ee}\sim 0$, but, in addition, is important to remark that the process is quite different from the standard one in which two left-handed electrons are produced. If $\nbb$ is found and proceeds as in the mechanism suggested in this paper, the produced electrons will be right-handed and, therefore, it will be possible, in principle, to distinguish this mechanism by measuring the polarization of the emitted electrons.

We have also found a DM candidate which can reproduce the observed relic abundance
yet can survive the current constraints from the direct detection experiments. 

Furthermore, our model provides the prospect of detecting new scalars with masses below $\order({\rm TeV})$ in
collider experiments (for LHC studies on lepton number violating singly and doubly charged scalars see for instance \cite{delAguila:2013yaa,delAguila:2013mia}). Among these new particles, $\chi^\pm$ and $\chi^{\pm\pm}$ being $Z_2$-odd,
cannot decay directly into the SM particles. A search strategy for these kinds of exotic charged scalars
can be interesting for the collider studies. Moreover, the decay branching ratios of the singlet doubly charged scalar $\kappa^{++}$ are controlled by the $f_{ab}$ couplings which are fixed in terms of the neutrino mass parameters, therefore, if $\kappa^{++}$ is found at the LHC it will be possible to distinguish this model from other models by comparing the $\kappa^{++}$ leptonic decay branching ratios to neutrino oscillation data and to LFV processes, which also depend on the same couplings.

%%%%%%%%%%%%%%%%%%%%%%%%%%%%%%%%%%%%%%%%%%%%%%%%%%%%%%%%
%================   Acknowledgements   ================================
%%%%%%%%%%%%%%%%%%%%%%%%%%%%%%%%%%%%%%%%%%%%%%%%%%%%%%%%
\section*{Acknowledgements} 
A.S. would like to acknowledge J.M. No for discussions. All Feynman diagrams have been drawn using JaxoDraw \cite{Binosi:2003yf,Binosi:2008ig}. This work has been partially supported by the Spanish MINECO under grants  FPA2011-23897, FPA2014-54459-P, by the ``Centro de Excelencia Severo Ochoa'' Programme under grant SEV-2014-0398 and by the ``Generalitat Valenciana'' grant GVPROMETEOII2014-087.

% % % % % % % % % % % % % % % % % % % % % % % % % % % % % % % % % % % % % % % % % %
%========================== Appendices ==============================
% % % % % % % % % % % % % % % % % % % % % % % % % % % % % % % % % % % % % % % % % %
\begin{appendices}
\numberwithin{equation}{section}
%

% % % % % % % % % % % % % % % % % % % % % % % % % % % % % % % % % % % % % % % % % % % %
% %============    kWW vertex     ==========================================
% % % % % % % % % % % % % % % % % % % % % % % % % % % % % % % % % % % % % % % % % % % %
\section{Computation of the loop induced $\kappa WW$ vertex}
\label{ap:Neutrinoless}
\begin{figure}[htbp!]
\centering
\includegraphics[scale=0.55]{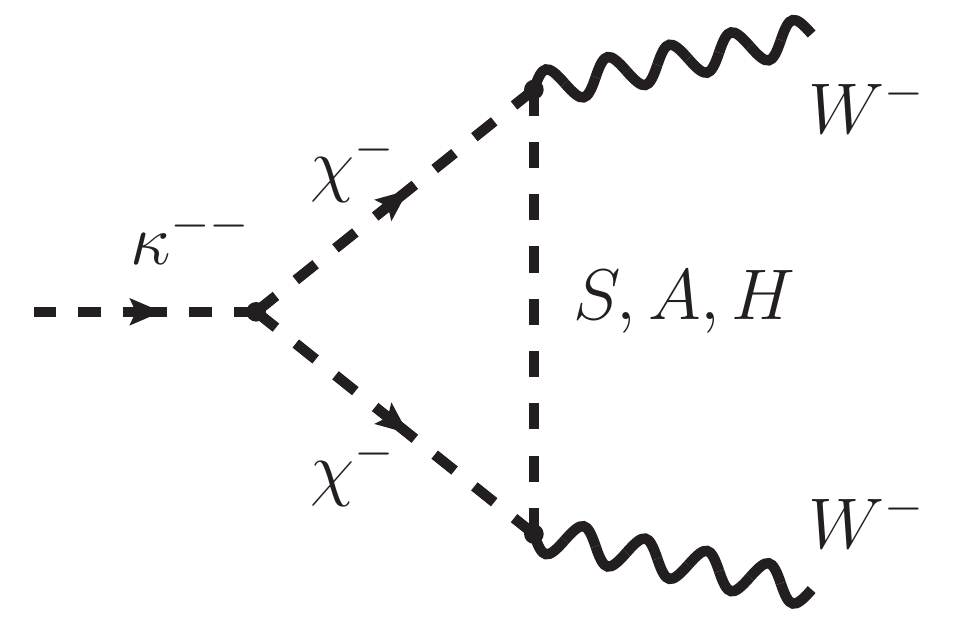} ~~ \includegraphics[scale=0.55]{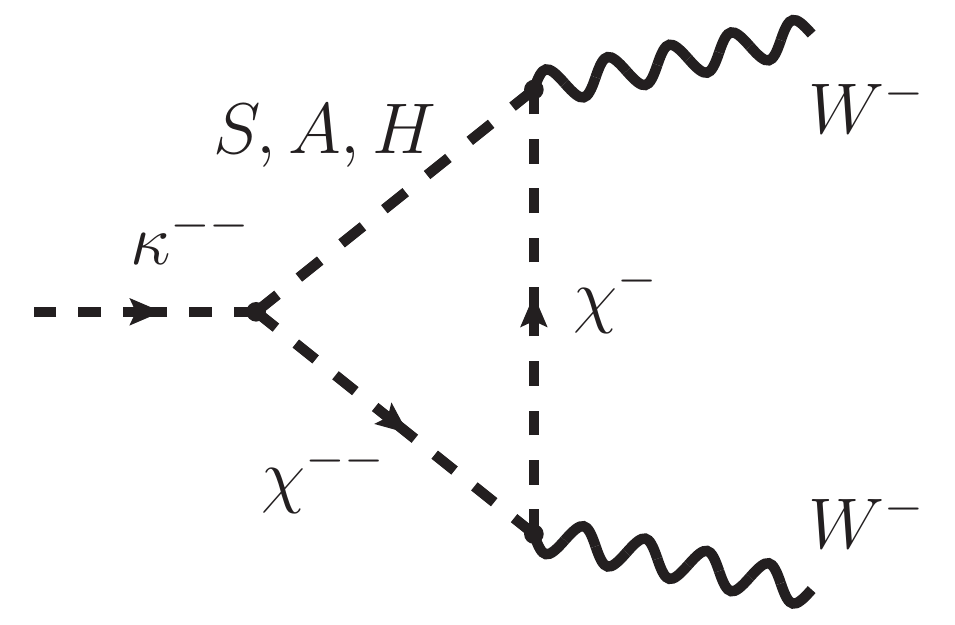}
~~ \includegraphics[scale=0.55]{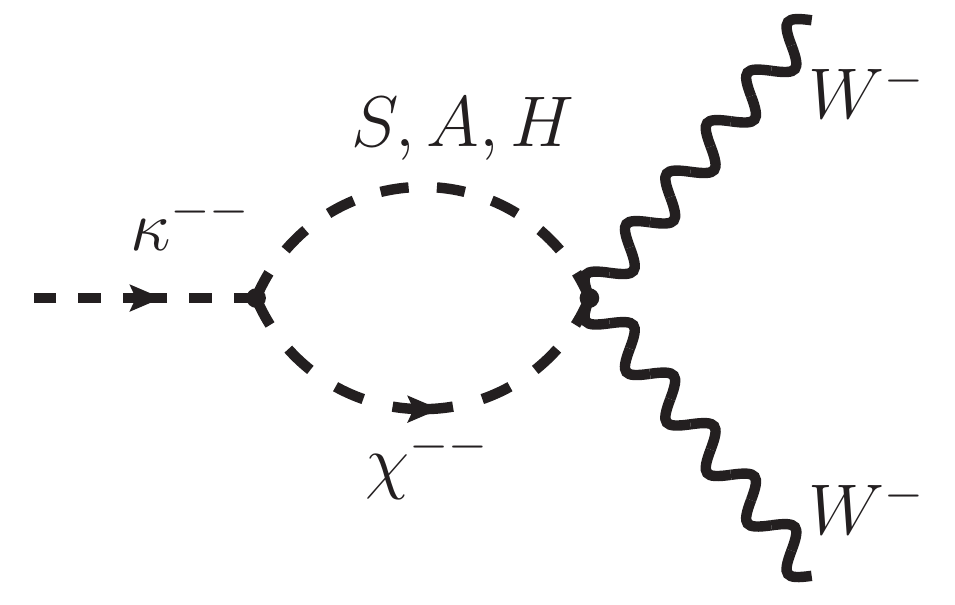}
\caption{One loop diagrams contributing to the $\kappa WW$ vertex in the unitary
gauge.
\label{f:kww}}
\end{figure}
Here we compute the effective $\kappa^{--}W_{\mu}^{+}W^{\mu+}$ vertex
at one loop for vanishing external momenta.
Our assumption is justified in view of the fact that the momentum transfers to $\kappa$ and $W$-bosons
in Fig.~\ref{fig:0nu2beta} are much smaller than the corresponding masses. We write the effective vertex
as\begin{eqnarray}
\label{e:kwwlag}
\ml_{\kappa WW} &=& C_{\kappa WW}\kappa^{--}W_{\mu}^{+}W^{\mu+} + {\rm h.c.} \,,
\end{eqnarray}
which, after spontaneous symmetry breaking, emerges from the following gauge invariant operator:
\begin{eqnarray}
\ml_{\kappa\mathrm{eff}}=C_{\kappa\mathrm{eff}}\kappa^{++}\left(\Phi^{\dagger}D^{\mu}\tilde{\Phi}\right) \left(\Phi^{\dagger}D_{\mu}\tilde{\Phi}\right)+\mathrm{h.c.}
\label{eq:effective-kappa}
\end{eqnarray}
After integrating out $\kappa^{++}$, \Eqn{eq:effective-kappa} leads to the following LFV gauge invariant operator\cite{delAguila:2012nu,delAguila:2011gr}:
\begin{eqnarray}
%\label{}
\ml_{eeWW}=C_{eeWW}\left(\overline{e_{R}}\ f_{ee}^{*}\ e_{R}^{c}\right)\left(\Phi^{\dagger}D^{\mu}\tilde{\Phi}\right)\left(\Phi^{\dagger}D_{\mu}\tilde{\Phi}\right) \,.
\end{eqnarray}

 We depict in Fig.~\ref{f:kww} the three diagrams that contribute to the vertex.
Each of these diagrams seem to diverge logaritmically. But one should keep in mind that
the neutral scalar exchange must violate lepton number conservation. Thus a large cancellation among the contributions
from the three neutral scalars, $A$, $H$ and $S$, is expected. After adding all the contributions 
we obtain  an effective neutral scalar propagator of the following form~(for
Minkowsky momenta)
\begin{equation}
\frac{1}{2}\frac{\sin^{2}\alpha\cos^{2}\alpha(m_{H}^{2}-m_S^{2})^{2}}{(p^{2}-m_{H}^{2})(p^{2}-m_S^{2})(p^{2}-m_{A}^{2})}=\frac{\lambda_{6}^{2}\langle\Phi\rangle^{4}}{(p^{2}-m_{H}^{2})(p^{2}-m_S^{2})(p^{2}-m_{A}^{2})}\,,
\label{eq:LNVCancellation}
\end{equation}
where, $\langle \Phi \rangle = v/\sqrt{2}$. Evidently, after adding contributions from $A$, $H$ and $S$,
every diagram in Fig.~\ref{f:kww} becomes finite individually. Now we can write the expression of
$C_{\kappa WW}$ (defined in \Eqn{e:kwwlag}) as follows:
\begin{eqnarray}
\label{e:ckww}
C_{\kappa WW} &=& \mu_{\kappa}g^{2}\lambda_{6}^{2}\langle\Phi\rangle^{4}\frac{1}{16\pi^{2}m_{A}^{4}}I_{\beta} \,,
\end{eqnarray}
with $I_{\beta}$ a function of the masses of the particles running
in the loop which contains three contributions corresponding to the three diagrams in Fig.~\ref{f:kww}.
Thus, we express  $I_{\beta}$ as follows:
\begin{eqnarray}
I_{\beta} &=& I_{\beta}^1 +I_{\beta}^2 +I_{\beta}^3 \,, ~~{\rm with,} \\
I_{\beta}^1 &=& m_{A}^{4}\int_{0}^{\infty}\dd{q} q^{3}\frac{q^{2}}{(q^{2}+m_{\chi^{+}}^{2})^{2}(q^{2}+m_{A}^{2})(q^{2}+m_{H}^{2})(q^{2}+m_S^{2})} \,, \\
I_{\beta}^2 &=& -2m_{A}^{4}\int_{0}^{\infty}\dd{q}  q^{3}\frac{1}{(q^{2}+m_{\chi^{++}}^{2})(q^{2}+m_{A}^{2})(q^{2}+m_{H}^{2})(q^{2}+m_S^{2})} \,, \\
I_{\beta}^3 &=& 2m_{A}^{4}\int_{0}^{\infty} \dd{q} q^{3}\frac{q^{2}}{(q^{2}+m_{\chi^{++}}^{2})(q^{2}+m_{\chi^{+}}^{2})(q^{2}+m_{A}^{2})(q^{2}+m_{H}^{2})(q^{2}+m_S^{2})} \,,
\end{eqnarray}
where we have passed to Euclidean momenta and integrated over the angular variables. Adding the three contributions we simplify the expression for $I_{\beta}$ as follows:
\begin{eqnarray}
I_{\beta} = m_{A}^{4}\int_{0}^{\infty}\dd{q} q^{3}\frac{q^{4}+q^{2}(m_{\chi^{++}}^{2}-2m_{\chi^{+}}^{2})-2m_{\chi^{+}}^{4}}{(q^{2}+m_{\chi^{++}}^{2})(q^{2}+m_{\chi^{+}}^{2})^{2}(q^{2}+m_{A}^{2})(q^{2}+m_{H}^{2})(q^{2}+m_S^{2})} \,.
\end{eqnarray}
We have checked that we obtain the same result
by using the equivalence theorem where the external $W$-bosons are replaced by 
the corresponding Goldstone bosons.

\begin{figure}[htbp!]
\centering
\includegraphics[scale=0.50]{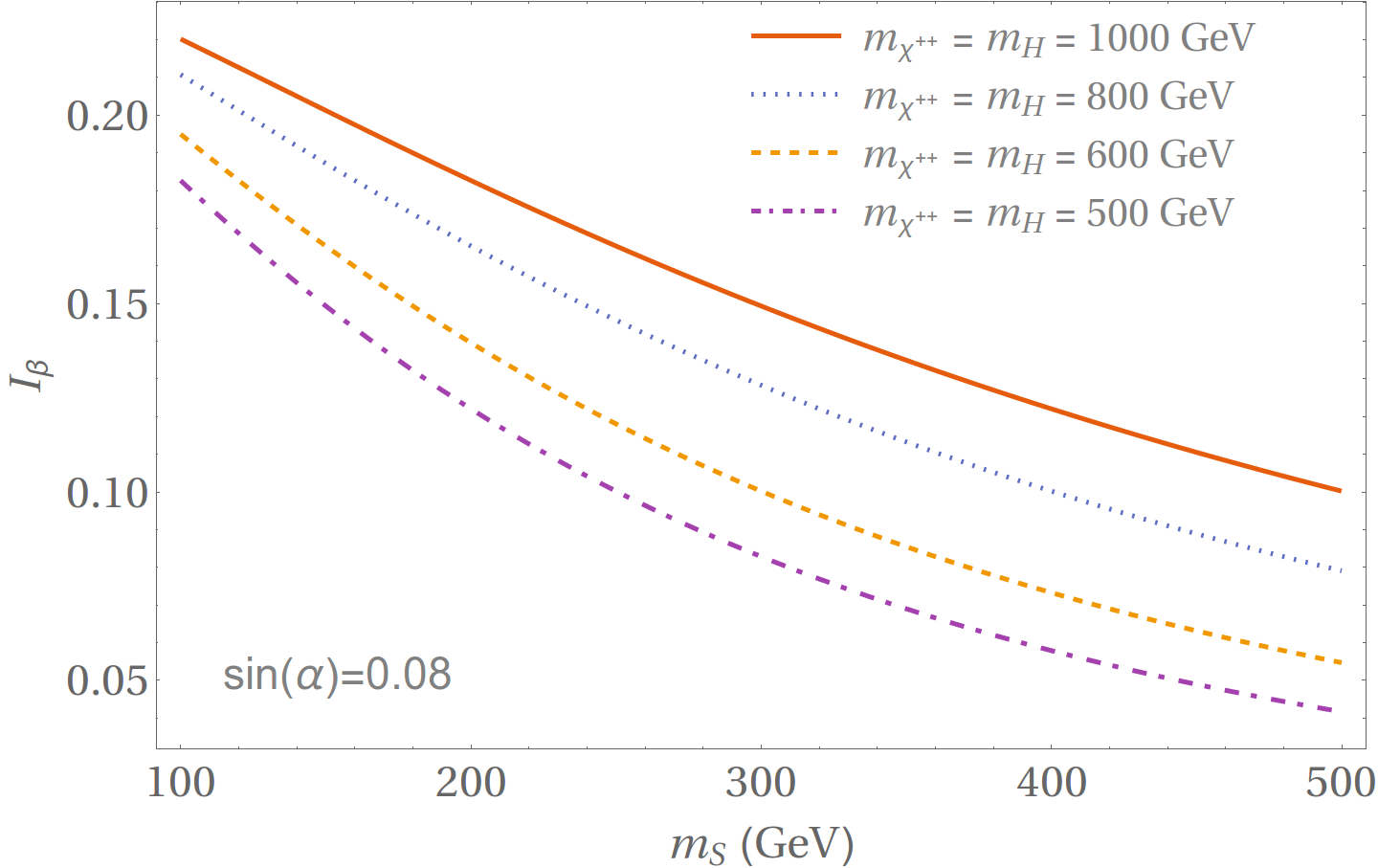}
\caption{The $0\nu\beta\beta$ integral, $I_\beta$, as a function of $m_S$ for some representative values of the other parameters. We fix $\sin(\alpha)=0.08$, use Eq.~(\ref{e:mA}) and Eq.~(\ref{e:corr}) and take $m_H=m_{\chi^{++}}$.}
\label{f:Ibeta}
\end{figure}

In the limit $m_{H}=m_{A}=m_{\chi^{++}}=m_{\chi^{+}}$ and 
$m_S\ll m_A$ we obtain $I_{\beta}\sim1/4$ while if all masses are equal we get $I_{\beta}=1/24$. If we fix $\sin(\alpha)$ $m_A$ can be obtained from $m_H$ and $m_S$ using Eq.~(\ref{e:mA}) while $m_{\chi^{+}}$ can be written in terms
of $m_{\chi^{++}}$ and $m_A$ using Eq.~(\ref{e:corr}). Thus, $I_\beta$ can be written as a function of $\sin(\alpha)$, $m_{\chi^{++}}$, $m_H$ and $m_S$ only. 
In Fig.~\ref{f:Ibeta} we present results for some representative values of the masses (we fix $\sin(\alpha)=0.08$ and give $I_\beta$ as a function of $m_S$ for different values of $m_H=m_{\chi^{++}}$).

%%%%%%%%%%%%%%%%%%%%%%%%%%%%%%%%%%%%%%%%%%%%%%%%%%%%%%%%%%%%%%%%%%%
%%%%%      Appendix: Neutrino mass computation            %%%%%%%%%%%%%%%%%%%%%%%%%%%%%%%%%%%
%%%%%%%%%%%%%%%%%%%%%%%%%%%%%%%%%%%%%%%%%%%%%%%%%%%%%%%%%%%%%%%%%%%
\section{Details of the calculation of the neutrino masses}
\label{ap:Neutrino-masses}
We define the Majorana mass matrix for the neutrinos as follows:
\begin{eqnarray}
%\label{}
\ml_{\rm majorana} =-\frac{1}{2}\overline{\nu_{L}^{c}}\cdot M\cdot \nu_{L}+\mathrm{h.c.}
\end{eqnarray}
Our parametrization for the elements of the neutrino mass matrix have been displayed in \Eqn{e:nuelements}
which, in terms of the physical parameters, can be rewritten as
\begin{eqnarray}
%\label{}
M_{ab}=\frac{8\mu_{\kappa}\sin^{2}2\alpha G_{F}^{2}(m_{H}^{2}-m_S^{2})^{2}}{(4\pi)^{6}\mkpp^{2}}I_\nu m_{a}f_{ab}m_{b} \,.
\end{eqnarray}
\begin{figure}
\begin{centering}
\includegraphics[scale=0.55]{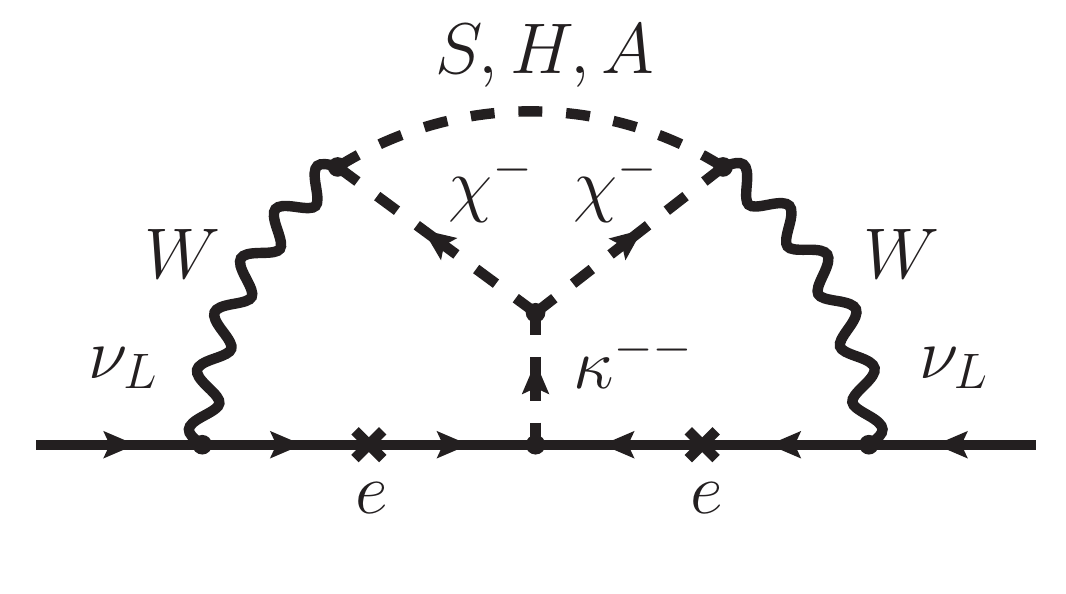} ~~~~ \includegraphics[scale=0.55]{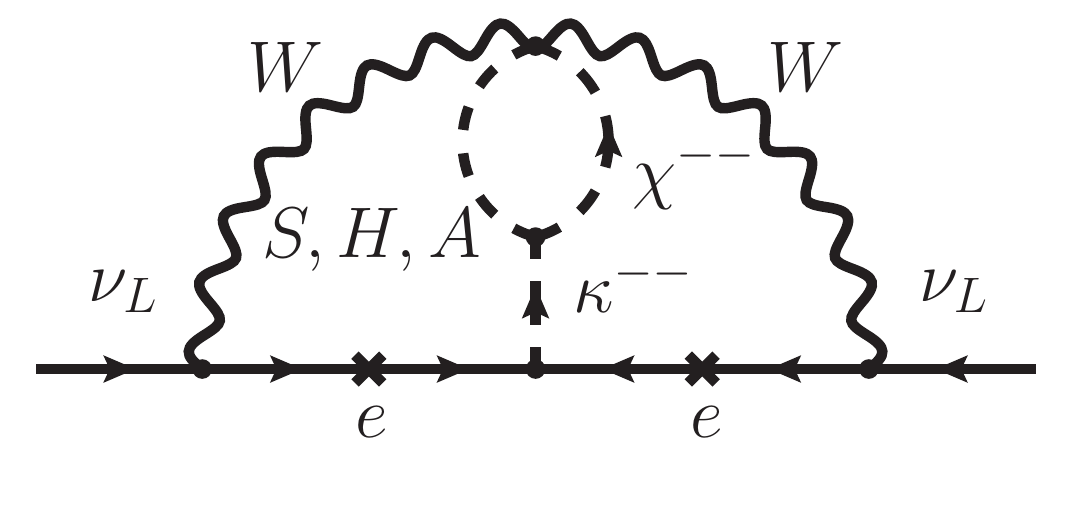}
\par\end{centering}
\begin{centering}
\includegraphics[scale=0.55]{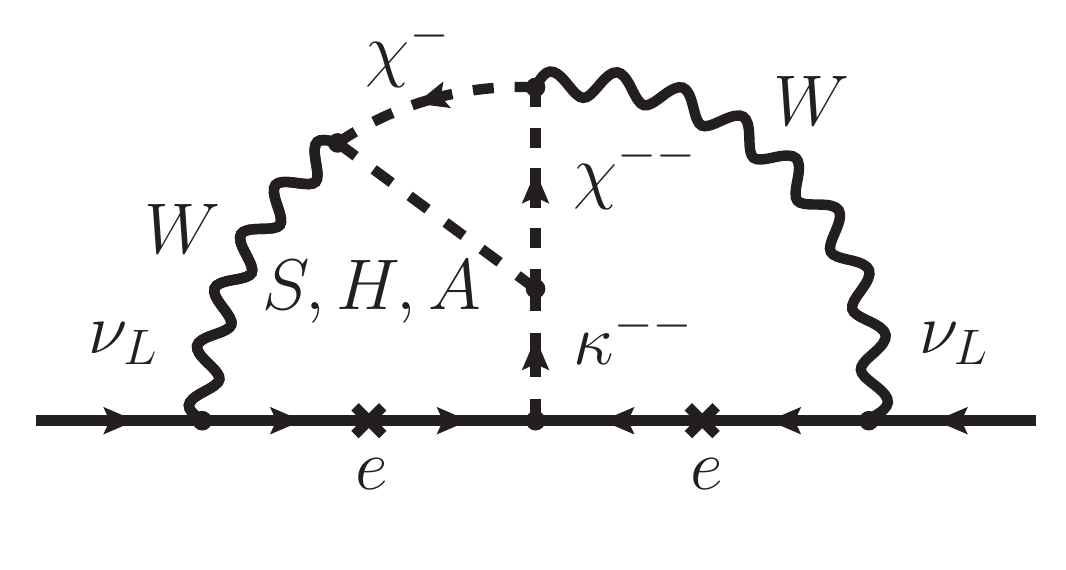} ~~~~ \includegraphics[scale=0.55]{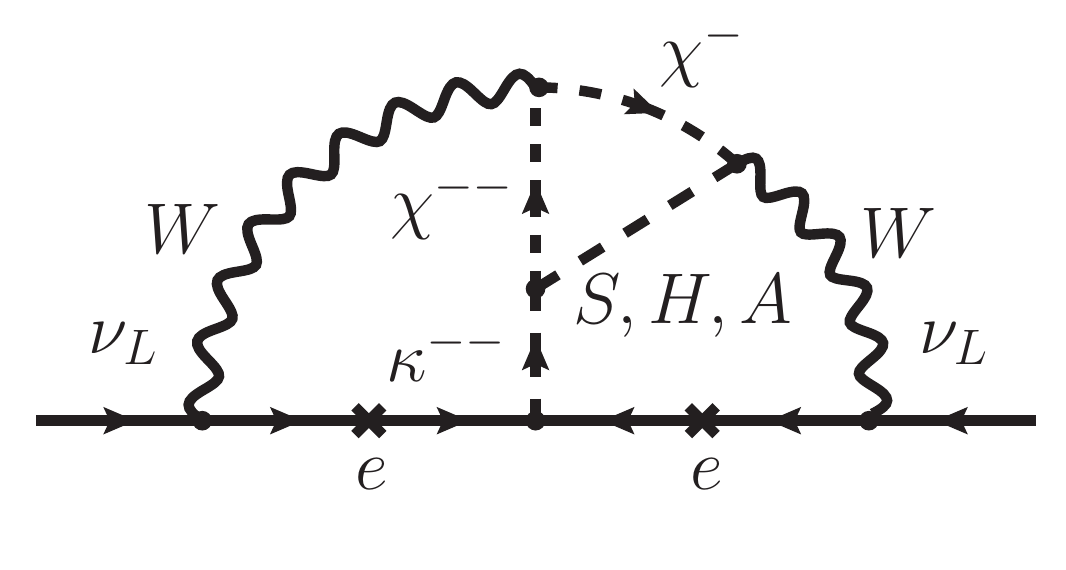}
\par\end{centering}
\caption{Three loop diagrams contributing to neutrino masses in the unitary
gauge.
\label{fig:numass-unitary}}
\end{figure}

In the unitary gauge there are four diagrams contributing to the neutrino masses as displayed in Fig.~\ref{fig:numass-unitary}.
As explained in Appendix~\ref{ap:Neutrinoless}, each diagram will be finite when we add together the contributions
from $H$, $S$ and $A$. Note that the two diagrams in the last row of Fig.~\ref{fig:numass-unitary}, after some relabeling
of momenta, will give identical contributions. Taking this into account, we decompose $I_\nu$ into three pieces as follows:
\begin{eqnarray}
\label{e:inu}
I_\nu = I_\nu^1 + I_\nu^2 + I_\nu^{34} \,.
\end{eqnarray}
Explicit expressions for the individual pieces in \Eqn{e:inu} are given below (all the momenta are Euclidean):
\begin{subequations}
\label{e:pieces}
\begin{eqnarray}
%\label{}
I_\nu^{1} &=& (4\pi)^{6}\mkpp^{2}\int_{q}P_{c}\frac{V_{1}\cdot V_{2}}{\left\{(q_{1}+q_{3})^{2}+m_{\chi^{+}}^{2}\right\}\left\{ (q_{3}-q_{2})^{2}+m_{\chi^{+}}^{2}\right\} } \,, \\
I_\nu^{2} &=& -2(4\pi)^{6}\mkpp^{2}\int_{q}P_{c}\frac{4M_{W}^{4}  +M_{W}^{2}(q_{1}^{2}+q_{2}^{2})+(q_{1}q_{2})^{2}}{\left\{(q_{3}+q_{1}+q_{2})^{2}+m_{\chi^{++}}^{2}\right\}} \,, \\
I_\nu^{34} &=& 2(4\pi)^{6}\mkpp^{2}\int_{q}P_{c}\frac{V_{1}\cdot V_{3}}{\left\{(q_{3}+q_{1}+q_{2})^{2}+m_{\chi^{++}}^{2} \right\}\left\{(q_{3}+q_{1})^{2}+m_{\chi^{+}}^{2}\right\} } \,,
\end{eqnarray}
\end{subequations}
\begin{subequations}
\label{e:defn}
\begin{eqnarray}
%\label{}
\hspace*{-10mm}{\rm with,} ~~~ P_{c}&=& \frac{1}{q_{1}^{2}(q_{1}^{2}+M_{W}^{2})q_{2}^{2}(q_{2}^{2}+M_{W}^{2}) \left\{(q_{1}+q_{2})^{2}+\mkpp^{2}\right\} (q_{3}^{2}+m_{H}^{2})(q_{3}^{2}+m_S^{2})(q_{3}^{2}+m_{A}^{2})}\,, \\
V_{1}^{\mu}&=& M_{W}^{2}(2q_{3}+q_{1})^{\mu}+\left\{(2q_{3}+q_{1})\cdot q_{1}\right\} q_{1}^{\mu} \,, \\
V_{2}^{\mu}&=& M_{W}^{2}\left(2q_{3}-q_{2}\right)^{\mu}+\left\{(2q_{3}-q_{2})\cdot q_{2}\right\} q_{2}^{\mu} \,, \\
V_{3}^\mu &=& M_{W}^{2}(2q_{3}+2q_{1}+q_{2})^{\mu} +\left\{(2q_{3}+2q_{1}+q_{2})\cdot q_{2}\right\}q_{2}^{\mu} \,.
\end{eqnarray}
\end{subequations}
To evaluate the integrals in \Eqn{e:pieces} we express the Euclidean four-momenta in the four dimensional
spherical polar coordinates as follows:
\begin{eqnarray}
\label{e:4vector}
q_{i}=q_{i}(\cos\psi_{i},~\sin\psi_{i}\text{\ensuremath{\cos\theta_{i}},~\ensuremath{\sin\psi_{i}\text{\ensuremath{\sin\theta_{i}\cos\phi_{i}},}~\sin\psi_{i}\text{\ensuremath{\sin\theta_{i}\sin\phi_{i}}}})} \,,
\end{eqnarray}
where, for brevity, we have used $q_i$ to denote both the four Euclidean vector and its modulus. With this, the differential
under the integral can be expressed as:
\begin{eqnarray}
%\label{}
\int_{q} \equiv \int\prod_{i=1}^{3}\frac{\dd{q_{i}} q_{i}^{3}}{(2\pi)^{4}} \dd{\phi_{i}} \dd{\theta_{i}} \sin\theta_{i} \dd{\psi_{i}} \sin^{2}\psi_{i}\:,\qquad\phi_{i}\in[0,2\pi]\,,\;\theta_{i}\in[0,\pi]\,,\;\psi_{i}\in[0,\pi] \,,\; q_{i}\in[0,\infty] \,.
\end{eqnarray}
Without any loss of generality we can orient our 1-axis in the direction of $q_3$ and express the momenta as follows:
\begin{eqnarray}
%\label{}
q_{3}=q_{3}(1,0,0,0)\,,~~ q_{2}=q_{2}(\cos\psi_{2},\, \sin\psi_{2},0,0)\,,~~ q_{1}=q_{1}(\cos\psi_{1},\, \sin\psi_{1}\text{\ensuremath{\cos\theta_{1}},~\ensuremath{\sin\psi_{1}\text{\ensuremath{\sin\theta_{1}},}}~0)} \,.
\end{eqnarray}
In this way, the integrands in \Eqn{e:pieces} will not depend on the angles $\phi_{1},\phi_{2},\theta_{2},\phi_{3},\theta_{3},\psi_{3}$
and they can be integrated out very easily. After this, the remaining six parameter integrals can be computed numerically (we have used
Mathematica along with the Cuba package for this purpose).  We have also checked numerically that, in the limit $g\to 0$ and small mixing, our unitary gauge calculation agrees with the calculation discussed in Sec. \ref{sec:nu-mass}, which includes only diagrams with scalar exchanges.

\begin{figure}[htbp!]
\centering
\includegraphics[scale=0.50]{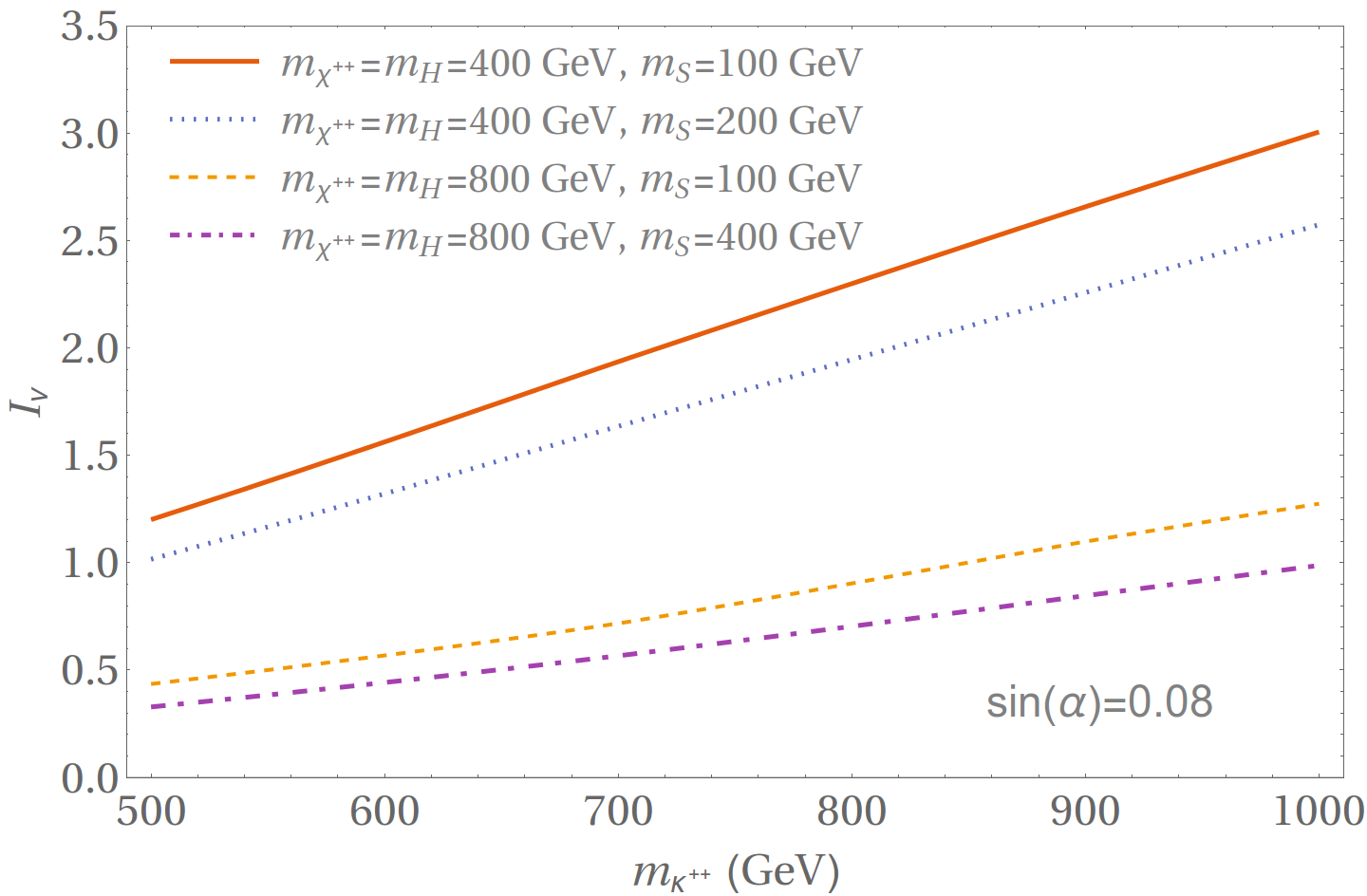}
\caption{The neutrino mass integral, $I_\nu$, as a function of $m_{\kappa^{++}}$ for some representative values of the other parameters. We fix $\sin(\alpha)=0.08$, use Eq.~(\ref{e:mA}) and Eq.~(\ref{e:corr}) and take $m_H=m_{\chi^{++}}$.}
\label{f:Inu}
\end{figure}

In Fig.~\ref{f:Inu} we give $I_\nu$ as a function of $m_\kappa$ for different values of the other parameters. As in Sec.~\ref{ap:Neutrinoless} we use Eq.~(\ref{e:mA}) and Eq.~(\ref{e:corr}), fix $\sin(\alpha)=0.08$ and take $m_H=m_{\chi^{++}}$).

\end{appendices}

%============================= References =====================================

\bibliographystyle{JHEP}
\bibliography{references.bib}
\end{document}